\documentclass[twocolumn,pra,longbibliography,superscriptaddress]{revtex4-2}
\usepackage{amsthm,amsmath,amssymb,amsfonts,graphicx,verbatim,xcolor,bm,physics} 
\usepackage{ulem,hyperref} 
\usepackage[utf8]{inputenc} 
\usepackage{siunitx} 

\newcommand{\beqn}{\begin{eqnarray}}
\newcommand{\eeqn}{\end{eqnarray}}

\newcommand{\br}{\bm r}
\newcommand{\bk}{\bm k}
\newcommand{\bq}{\bm q}

\newcommand{\bb}{\bm b}
\newcommand{\ba}{\bm a}

\begin{document}

\title{Chern mosaic and ideal flat bands in equal-twist trilayer graphene}

\author{Daniele Guerci} 
\affiliation{Center for Computational Quantum Physics, Flatiron Institute, New York, New York 10010, USA}
\author{Yuncheng Mao} 
\affiliation{Universit\'e Paris Cit\'e, CNRS,  Laboratoire  Mat\'eriaux  et  Ph\'enom\`enes  Quantiques, 75013  Paris,  France}
\author{Christophe Mora} 
\affiliation{Universit\'e Paris Cit\'e, CNRS,  Laboratoire  Mat\'eriaux  et  Ph\'enom\`enes  Quantiques, 75013  Paris,  France}

\begin{abstract}
    We study trilayer graphene arranged in a staircase stacking configuration with equal consecutive twist angle. On top of the moiré cristalline pattern, a supermoiré long-wavelength modulation emerges that we treat adiabatically. For each valley, we find that the two central bands are topological with Chern numbers $C=\pm 1$ forming a Chern mosaic at the supermoiré scale. 
    The Chern domains are centered around the high-symmetry stacking points ABA or BAB and they are separated by gapless lines connecting the AAA points, where the spectrum is fully connected. In the chiral limit and at a magic angle of $\theta \sim 1.69^\circ$, we prove that the central bands are exactly flat with ideal quantum curvature at ABA and BAB. Furthermore, we decompose them analytically as a superposition of an intrinsic color-entangled state with $\pm 2$ and a Landau level state with Chern number $\mp 1$. 
    To connect with experimental configurations, we also explore the non-chiral limit with finite corrugation and find that the topological Chern mosaic pattern is  indeed robust and the central bands are still well separated from remote bands.
\end{abstract}

\maketitle

\begin{figure}
    \centering
    \includegraphics[width=0.95\linewidth]{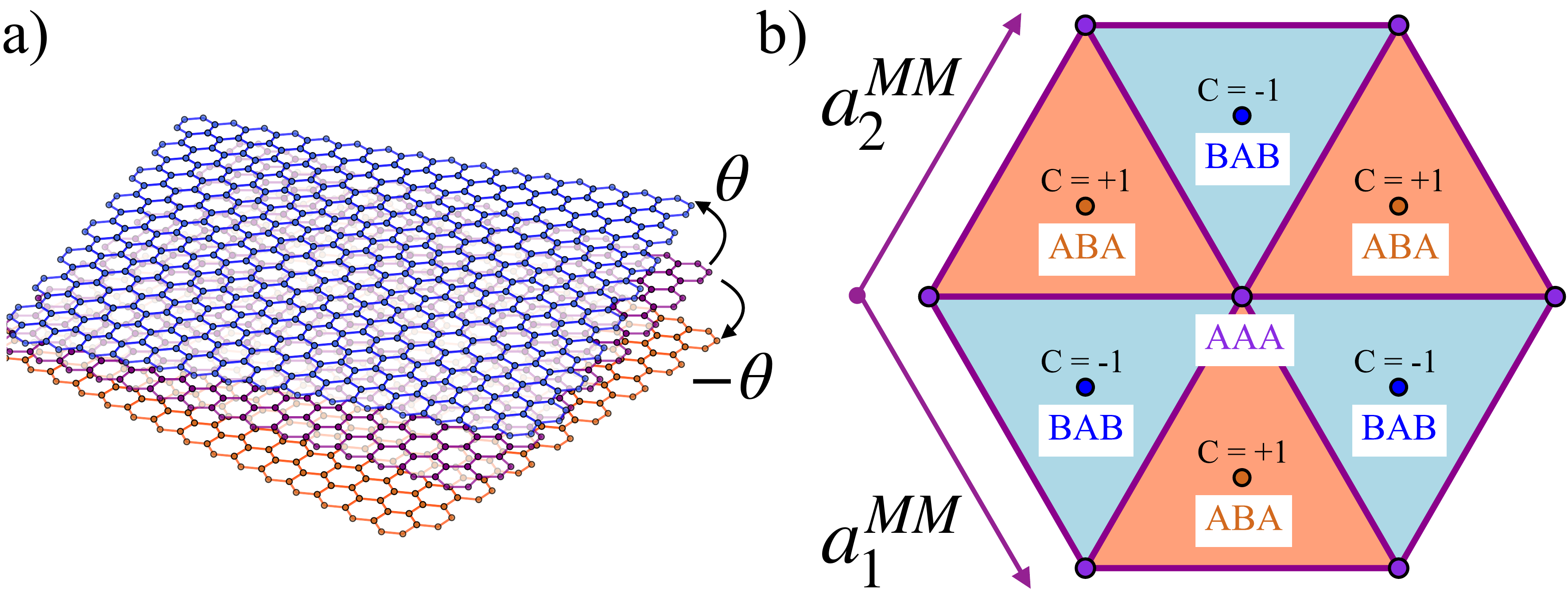}
    \caption{a) Equal-twist angle trilayer graphene lattice in real space. b) Real-space Chern mosaic in the moir\'e of moir\'e lattice scale. We distinguish three high-symmetry regions of character AAA, ABA and BAB. The former shows a fully connected spectrum while the latter is characterized by flat bands with total Chern number $+1$ and $-1$, respectively. Domain walls regions marking topological transition between Chern $\pm1$ states show fully connected spectrum.}
    \label{fig:chern_mosaic}
\end{figure}

\paragraph*{Introduction ---} 
Stacking and twisting two layers of graphene realizes an extraordinary platform~\cite{Andrei:2020aa} which in the magic angle region gives rise to flat bands~\cite{Bistritzer12233,bistritzer2011moirebutterfly,Santos,Santos2,Morell,Mele_2010} hosting superconductivity~\cite{Cao:2018aa,Cory_Science19,Lu_2019, stepanov2020untying,Young_Naturephys20}, interaction-driven insulating states~\cite{Cao:2018ab,tomarken2019electronic,Young_Naturephys19,wong2020cascade,zondiner2020cascade,Cao2020Strange,park2021flavour,jaoui2022quantum,Bultinck_2020_PRX,lian2020tbg}, anomalous Hall effects~\cite{Sharpe605,Serlin900,pixley2019ferromagnetism,Chen:2020aa,Li2020Experimental,Grover_2022,Zaletel_TBG_AQH,Cecile_PRL20,he2020giant,Potasz2021Exact} and fractional Chern insulators~\cite{Xie_2021_FCI,Repellin_fractional_2020,parker2021fieldtuned}. The intimate connection between the flat bands of twisted bilayer graphene (TBG) and the properties of Landau levels~\cite{Tarnopolsky2019,XiDai_PseudoLandaulevel,popov2020hidden,Ledwith_2020,Bernevig_tbg1,bernevig2020tbg,Eslam_hierarchy,ZhaoLiu_TBG,RafeiRen_TBG,Khalaf_2021,Ledwith_ann_2021,Wang_2021,Sheffer_2021,hFL_Bernevig_2022,Stern_2023} played a key role for the understanding of the interplay between correlation and topology in the aforementioned correlated states. Following this guiding principle we characterized the properties of the flat bands in equal-twist staircase trilayer graphene (eTTG), see Fig.~\ref{fig:chern_mosaic}a, finding high-symmetry stacking ABA/BAB configurations with total Chern number $\pm1$ hosting an intrinsic color-entangled state~\cite{Barkeshli_2012,Yangle_Modelwf,YangLe_2014,wang2022origin} with Chern number $2$ and a Landau level like state with Chern number $-1$.    

Adding an additional graphene sheet to TBG rotated by a small relative twist angle [twisted trilayer graphene (TTG)] gives rise to the superposition of two moir\'e superlattices~\cite{Zhang2021Correlated,uri2023superconductivity}. With the exception of mirror-symmetric TTG~\cite{park2021tunable,cao2021pauli,Hao_2021,kim2022evidence,liu2022isospin,Carr_2020,Caligaru_2021,Xie_2021,Daniele_2022_mTTG,Christos_PRX_2022}, the two moir\'e periodicities are incommensurate~\cite{PhysRevLett.125.116404,Yuncheng2023} leading to a quasicristalline structure that dominates the electronic behavior at relevant energies~\cite{uri2023superconductivity}.  
The theoretical description of twisted trilayer graphene runs into fundamental difficulties~\cite{PhysRevLett.125.116404,Yuncheng2023} due to the quasiperiodic nature of the low-energy Hamiltonian, disallowing all the simplifications from Bloch's theorem. Similar effects can also emerge in encapsulated TBG when the hBN layers are nearly aligned with the moir\'e pattern of TBG~\cite{Cea_2020,Shi_2021}. 

The aim of this letter is to study the emergent effect of the superposition of the two moir\'e patterns in eTTG, see Fig.~\ref{fig:chern_mosaic}a. The system is the simplest example of a quasiperiodic moir\'e crystal~\cite{PhysRevLett.125.116404} where the angle $\theta_{12}$, between layer 1 (top) and 2 (middle), and $\theta_{23}$, between layer 2 and 3 (bottom), are equal $\theta_{12}=\theta_{23}\equiv\theta$. In the magic angle region, where $\theta\approx1^\circ$, the two incommensurate periodicity can be decomposed in a fast modulation $\bq_j$ on the moir\'e scale $|\bq_j|\propto\theta$ and a slow one $\delta\bq_j$ with much larger periodicity $|\delta\bq_j|\propto\theta^2$~\cite{Yuncheng2023,Suppmat}. Applying semiclassical adiabatic approximation~\cite{Kohn_Luttinger_1955,Bastard_1981,White_1981,Bastard_1982,RevModPhys.62.173} we define a local  Hamiltonian $H_{\rm eTTG}(\br)=H_{\rm eTTG}(\br,{\bm \phi})$ where ${\bm \phi}$ depends on the slowly varying supermoir\'e scale~\cite{Yuncheng2023}. In this picture, we obtain a Chern number versus ${\bm \phi}$ real-space map Fig.~\ref{fig:chern_mosaic}b that gives rise to a triangular lattice Chern mosaic of regions with $\pm1$ Chern number. The domain walls separating the topological regions close the gaps to the remote bands and form lines connecting the AAA centers.

There are three high-symmetry stacking configurations that are especially significant and indicative of the Chern mosaic pattern: AAA, ABA and BAB. We explore them analytically in the chiral limit to unveil the topological features of the mosaic. 
We thereby derive analytical expressions for the resulting ideal flat bands emerging at a magic angle.
The AAA stacking, considered in the preprint~\cite{popov2023magic}, is characterized by a vanishing Berry curvature and a fully connected spectrum protected by $C_{2z}T$~\cite{Christophe_2019}. At the magic angle $\theta_{AAA}\approx0.75^\circ$ a fourfold degenerate zero energy flat band sector emerges, connected to a single Dirac cone.
The ABA(BAB) stacking, on the other hand, shows a flat band region detached from the remote bands with total Chern number $C=1(-1)$. The origin of the finite Chern number is readily traced out by the nature of the flat bands at the magic angle $\theta_{ABA}\approx1.69^\circ$ which, remarkably, is larger than the one in mirror-symmetric TTG~\cite{Khalaf_2019}. We prove that the flat band sector decomposes into a Chern $+2(-2)$ color-entangled zero mode~\cite{wang2022origin,ledwith2022vortexability,mera2023uniqueness} and a Chern $+1(-1)$ Landau level like state~\cite{Tarnopolsky2019,Ledwith_2020,popov2020hidden,Wang_2021,JieWang_exactlldescription,Sheffer_2021,Stern_2023,parhizkar2023generic}. The resulting imbalance in Chern flux creates a Chern mosaic pattern in real space, which could potentially be detected by measuring the local orbital magnetization in real space~\cite{Li2020Experimental,Grover_2022}.

\begin{figure}
    \centering
    \includegraphics[width=0.4\textwidth]{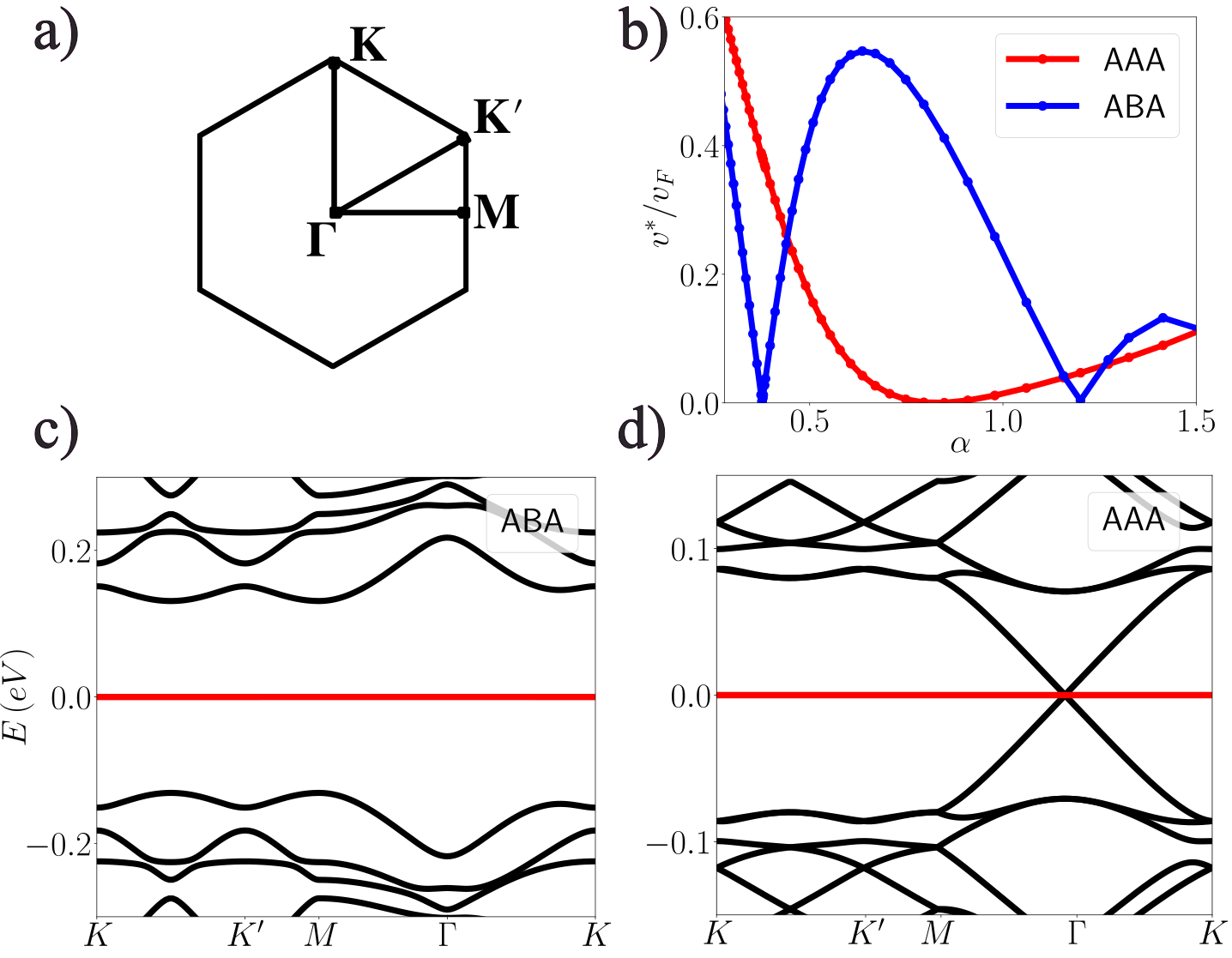}
    \caption{a) mini Brillouin zone (BZ). b) Renormalized velocity $v^*$ for AAA (red) and ABA (blue) regions as a function of the dimensionless coupling $\alpha=w_{AB}/v_Fk_\theta$ and $w_{AA}=0$ (chiral limit). Dispersion relation for ABA stacking panel c) and AAA stacking panel d) at the magic angle $\theta_{\rm AAA}\approx0.75^\circ$ and $\theta_{\rm ABA}\approx1.69^\circ$, respectively.}
    \label{fig:magic_flat_bands}
\end{figure}

\paragraph*{Chern Mosaic on the supermoir\'e scale ---} When the twisting angle is small, non commensurability effects are characterized by a length scale well separated from the moir\'e scale, $|\delta \bq_j|/|\bq_j|\approx0.02$ for $\theta=1^\circ$. As a result the long wavelength modulation can be treated parametrically, leading to the local Hamiltonian obtained in Ref.~\cite{Yuncheng2023}: 
\begin{equation}
\label{H_trilayer_staircase}
    H_{\rm eTTG}(\br,{\bm \phi}) =\begin{bmatrix}
    v_F\hat{\bk}\cdot\bm \sigma & T(\br,{\bm \phi}) & 0\\
    h.c. & v_F\hat{\bk}\cdot\bm \sigma & T(\br,-{\bm \phi}) \\
    0 & h.c. & v_F\hat{\bk}\cdot\bm \sigma
    \end{bmatrix},
\end{equation}
where $v_F\approx 10^6$m/s is the graphene velocity, the phase ${\bm \phi}=(\phi_1,\phi_2,\phi_3)$ defines the local stacking configuration~\cite{Yuncheng2023,Suppmat}. Varying ${\bm \phi}$ maps out the supermoiré unit cell in Fig.~\ref{fig:chern_mosaic}b, ${\bm \sigma}$ is the vector of Pauli matrices in the sublattice space and $\hat\bk=-i\nabla_{\br}$. The tunneling between different layers is described by the moir\'e potential: 
\begin{equation}
    T(\br,{\bm \phi})=\sum^{3}_{j=1} T_j e^{-i\br\cdot\bq_j}e^{-i\phi_j},
\end{equation}
where $T_{j+1} = w_{\rm AA} \sigma^0 + w_{\rm AB}[\sigma^x\cos2 \pi j/3  + \sigma^y\sin 2 \pi j/3]$, $w_{\rm AB}=110$meV, using complex notation $\bq_{j+1}= ie^{2i\pi j/3}$~\cite{Christophe_2019}, $j=1,2,3$, in unit of $k_\theta=\theta K_D$ with $K_D=4\pi/3a_{\rm G}$ and $a_{\rm G}\approx2.46$\AA. The moir\'e lattice is characterized by the reciprocal lattice vectors $\bb_{1/2}=\bq_1-\bq_{2/3}$ and primitive vectors $\ba_{1/2}$. 
Bloch periodicity takes the form $H_{\rm eTTG}(\br+\ba_{1/2},{\bm \phi})=U_{\varphi/-\varphi} H_{\rm eTTG}(\br,{\bm \phi}) U^\dagger_{\varphi/-\varphi}$ with $\varphi=2\pi/3$ and $U_\varphi=\text{diag}(\omega^*,1,\omega)$ with $\omega=e^{2\pi i/3}$. The spectrum is thus invariant upon shifting $\bk$ by $\ba_1$ and $\ba_2$ up to a layer dependent phase factor. The Hamiltonian~\eqref{H_trilayer_staircase} is also invariant under the particle-hole transformation $ P H_{\rm eTTG}(\br,{\bm \phi}) P = - H_{\rm eTTG}({-\br,\bm \phi})$ where
\begin{equation}
\label{particle_hole_symmetry}
    P= \begin{bmatrix}
        0 & 0 & 1 \\
        0 & -1 & 0 \\
        1 & 0 & 0
    \end{bmatrix}_{\rm layer}\otimes \sigma^0\equiv \mathcal{M}\otimes \sigma^0.
\end{equation}
At low-energy Eq.~\ref{H_trilayer_staircase} is characterized by three inequivalent Dirac cones at $K$, $K'$ and $\Gamma$ of the mini Brillouin zone (BZ) shown in Fig.~\ref{fig:magic_flat_bands}a. The central one at $\Gamma$ is protected by $P$ while $K$ and $K'$ are gapped for generic ${\bm \phi}$~\cite{Yuncheng2023}. 

We now obtain the spectrum of the Hamiltonian~\eqref{H_trilayer_staircase} and study the topological properties of the nearly-flat bands around charge neutrality. Fig.~\ref{fig:chern_mosaic}b shows the real-space mosaic pattern obtained by computing the Chern number $\mathcal C({\bm \phi})$ for the two central bands at the magic angle $\theta_{\rm ABA} = 1.69^\circ$, for finite corrugation $w_{\rm AA} = 0.8 w_{\rm AB}$.
The mosaic exhibits a triangular periodic structure, which is generated by the lattice vectors $\mathbf{a}_{1/2}^{\mathrm{MM}} = 4\pi e^{\mp i\pi/3}/3k{\theta}^{\mathrm{MM}}$, where $k_{\theta}^{\mathrm{MM}} = \theta^2 K_D$. The two central bands are topological everywhere except along lines connecting the AAA centers, where the spectrum is fully connected. Each topological region is centered around $\mathbf{r}_{\mathrm{ABA}} = (\mathbf{a}_{1}^{\mathrm{MM}} - \mathbf{a}_{2}^{\mathrm{MM}})/3$ and $\mathbf{r}_{\mathrm{BAB}} = -\mathbf{r}_{\mathrm{ABA}}$, with opposite Chern numbers of $\pm1$. Therefore, we specifically focus on these two high-symmetry points and consider the chiral limit, where the bands become exactly flat, and an analytical solution can be obtained.


\paragraph*{ABA-stacking: color-entangled flat band---} 
The ABA region is described by the local Hamiltonian $ \mathcal {H}_{\rm ABA}=H_{\rm eTTG}({\bm \phi}_{\rm ABA}) $ Eq.~\ref{H_trilayer_staircase} with ${\bm\phi}_{\rm ABA}=(0,\varphi,-\varphi)$. 
Here, the $C_{3z}$ symmetry is recovered while $C_{2x}$ and $C_{2z}T$ are broken. The latter connects ${\rm ABA}$ to ${\rm BAB}$, $C_{2z}T{\mathcal H_{\rm ABA}}(\br)(C_{2z}T)^\dagger=-{\mathcal H_{\rm BAB}}(-\br)$, explaining the opposite Chern number of the ABA and BAB regions. The combination of $C_{2x}$ and $C_{2z}T$ is a symmetry for the model $C_{2y}T$ which together with $P$ and $C_{3z}$ protects three Dirac cones at $\Gamma$, $K$ and $K'$~\cite{Suppmat}. We henceforth consider the chiral limit $w_{\rm AA}=0$ where an inspiring mathematical structure emerges~\cite{Tarnopolsky2019}.
$\mathcal {H}_{\rm ABA}$ then anticommutes with the chiral operator $\Lambda^z=\tau^0\otimes\sigma^z$ with $\tau^0$ the identity in the layer basis. Denoting with $\psi_l$ and $\chi_l$ with $l=1,2,3$ the wavefunction components polarized in the A and B sublattices, in the basis $\Psi=\begin{pmatrix} \psi_1 & \psi_2 & \psi_3 & \chi_1 & \chi_2 & \chi_3 \end{pmatrix}^T$ the Hamiltonian $\mathcal H_{\rm ABA}$  reads: 
\begin{equation}
\label{H_ABA}
  \frac{\mathcal H_{\rm ABA}(\br)}{v_Fk_\theta}=\begin{pmatrix}
    0 & \mathcal D_1(\br)\\
   \mathcal D^\dagger_1(\br) & 0 
    \end{pmatrix},  
\end{equation}
we look for zero mode solutions: 
\begin{equation}
    \mathcal D_1(\br){\bm\chi}_{\bk}(\br)=0,\quad \mathcal D^\dagger_1(\br){\bm\psi}_{\bk}(\br)=0,
\end{equation}
with boundary conditions: 
\begin{equation}\label{bloch-periodic}
\begin{split}
    &\bm\psi_{\bk}(\br +\bm a_{1/2})=e^{i\bk\cdot\bm a_{1/2}}U_\varphi \bm\psi_{\bk}(\br),\\
&\bm\chi_{\bk}(\br +\bm a_{1/2})=e^{i\bk\cdot\bm a_{1/2}}U_\varphi \bm\chi_{\bk}(\br).
\end{split}
\end{equation}
In Eq.~\ref{H_ABA} the operator is: 
\begin{equation}
\label{differential_operator_D_1}
    \mathcal D_1(\br)=\begin{pmatrix}
    -i\sqrt{2}\partial & U_\omega(\br) & 0 \\
    U_0(-\br) & -i\sqrt{2}\partial & U_0(\br)\\
   0 & U_\omega(-\br) & -i\sqrt{2}\partial
    \end{pmatrix},
\end{equation}
with $\partial=\partial_x-i\partial_y/(\sqrt{2}k_\theta)$, $U_0(\br)=\alpha\sum^{3}_{j=1} e^{-i\bq_j\cdot\br}$ and $U_\omega(\br)=\alpha\sum^{3}_{j=1} \omega^{j-1}e^{-i\bq_j\cdot\br}$ with $\alpha=w_{\rm AB}/v_Fk_\theta$. We also introduce $\bar\partial=\partial_x+i\partial_y/(\sqrt{2}k_\theta)$, $z=k_\theta (x+iy)/\sqrt{2}$ and $\bar z=k_\theta( x-iy)/\sqrt{2}$. 
We focus on the first magic angle $\theta_{ABA}\approx1.69^\circ$ where the renormalized velocity $v^*$ vanishes, see blue line in Fig.~\ref{fig:magic_flat_bands}b, and correspondingly the bands around charge neutrality becomes perfectly flat as shown in  Fig.~\ref{fig:magic_flat_bands}c. Interestingly, the single particle gap that separates the flat bands from remote ones is $E_{\rm gap}\approx130$meV quite large if compared with the typical value of the Coulomb interaction screened by metallic gates~\cite{bernevig2020tbg}.  
The $C_{3z}$ symmetry yields $\chi_{\Gamma1}(0)=\chi_{\Gamma3}(0)=0$, while $\chi_{\Gamma2}(0)$ is usually non-zero. The magic angle $\theta_{\rm ABA}$ is exactly defined by $\chi_{\Gamma,2} (0)=0$. As the spinor $\bm \chi_{\Gamma}(0)$ then fully vanishes at $\theta_{\rm ABA}$, the B-polarized flat band has an analytical expression
\begin{equation}
\label{flat-band_B}
    \bm\chi_{\bk}(\br)=\bar\eta_{\bk}(\bar z)\bm\chi_\Gamma(\br),
\end{equation}
where the antiholomorphic $\bar\eta_{\bk} (\bar z) = \eta_{\bk}^* (-z)$ is related to the meromorphic function
\begin{equation}\label{holomorphic}
   \eta_{\bk} (z) = e^{i k_1 z/a_1} \frac{\vartheta_1[ z/a_1 - k/b_2,\omega]}{\vartheta_1( z/a_1,\omega)},
\end{equation}
with the notation $k_1= \bk \cdot \ba_1$ and $\vartheta_1( z , \omega)$ is the Jacobi theta-function~\cite{Suppmat}, which vanishes at $z=0$ and results in a Bloch periodicity Eq.~\eqref{bloch-periodic}. The Bloch wavefunction associated with Eq.~\eqref{flat-band_B} is $k$-antiholomorphic
\begin{equation}
\begin{split}
    {\bm u}_{\bar k} (\br) =e^{-i\br\cdot{\bm b}_2\bar k/b^*_2}\frac{\vartheta_1[ {\bar z}/a_1^* + {\bar k}/b_2^*,-\omega^*]}{\vartheta_1[ {\bar z}/a_1^*,-\omega^*]}{\bm \chi}_\Gamma(\br)
\end{split}
\end{equation}
corresponding to an ideal flat band~\cite{Ledwith_2020,Wang_2021}. The Chern number of the band can be readily read off from the $k$-space boundary conditions 
\begin{equation}
\label{k_space_BC}
{\bm u}_{\bar k+\bar b_j}( \bk)= e^{-i{\bm b}_j\cdot \br }e^{i\phi_{{\bar k,\bar b_j}}} {\bm u}_{\bar k}(\br), 
\end{equation}
where $\phi_{{\bar k,\bar b_1}}=-2\pi\bar k/b^*_2+\pi-\pi b^*_1/b^*_2$ and $\phi_{{\bar k,\bar b_1}}=\pi$ which implies $C_B=-1$ where the Chern number has been computing employing~\cite{Wang_2021,wang2022origin}:
\begin{equation}
    C=\frac{\phi_{\bk_0+\bm b_2,\bm b_1}+\phi_{\bk_0,\bm b_2}-\phi_{\bk_0,\bm b_1}-\phi_{\bk_0+\bm b_1,\bm b_2}}{2\pi}.
\end{equation}

\begin{figure}
    \centering
    \includegraphics[width=0.45\textwidth]{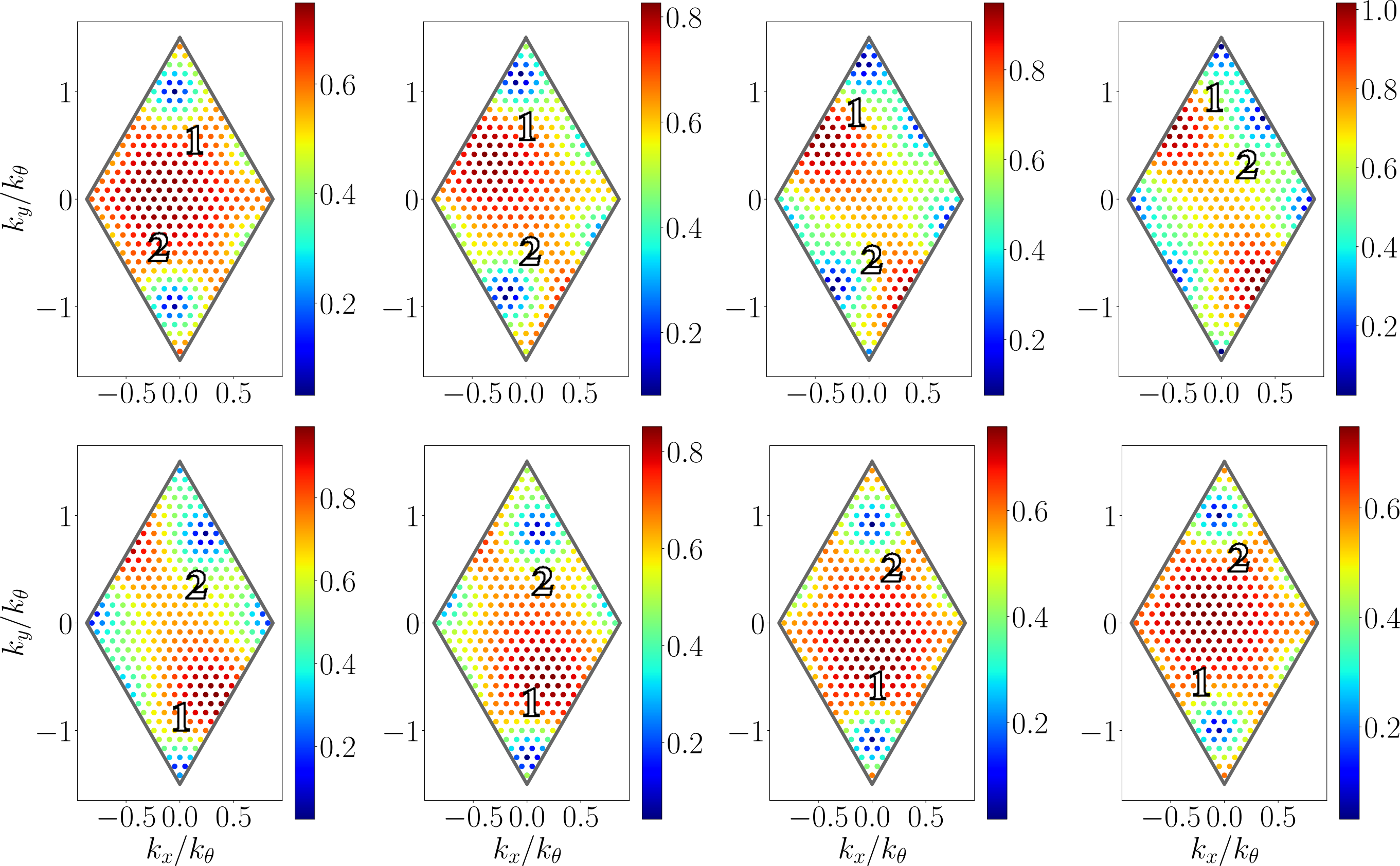}
    \caption{Each figure shows $|\psi_{\bk 2}(\br)|$ as a function of $\bk$ for a fixed $\br$. From left top to right bottom the position $\br$ evolves from $\br_0=(0.1,0.3)$ (unit of $1/k_\theta$) to $\br_0+\ba_1$. There are in total $C_A=2$ zeros in the Brillouin zone at fixed position $\br$, and their pattern is lattice translational invariant but indices are be exchanged going from $\br_0$ to $\br_0+\ba_1$. The effect resembles Thouless charge pumping but in reciprocal space. The solid gray line denotes the diamond shaped BZ. } \label{fig:nodes_swapping}
\end{figure}

We turn to the exact solution for the A-polarized wavefunction ${\bm \psi}$. $C_{3z}$ yields again $\psi_{\Gamma 1/3}(0)=0$ at $\Gamma$ and $\psi_{K/K' 2}(0)=0$ at $K$ ($K'$), however $\psi_{\Gamma 2}(0)$ does not vanish at the magic angle. Nevertheless, we numerically find that
$   \psi_{K,1}(0) = -  \psi_{K,3}(0)$,
right at the magic angle  $\theta_{\rm ABA}$ which, combined with particle-hole symmetry $P$, can be used to prove that the two spinors 
\begin{equation}\label{equal-spinor}
\psi_{K} (0 )= \psi_{K'} (0)    
\end{equation}
at $\theta_{\rm ABA}$. This remarkable identity allows us to exhibit an exact analytical expression for the A-polarized flat band (up to $\bk$-dependent prefactor)
\begin{equation}\label{flatAband}
   {\bm \psi}_{\bk}(\br) =   a_{k} 
   \eta_{\bk - \bq_1} (z) {\bm \psi}_{K}(\br) +  a_{-k} \eta_{\bk + \bq_1} (z) {\bm \psi}_{K'}(\br),
\end{equation}
satisfying the Bloch periodicity Eq.~\eqref{bloch-periodic},
with the holomorphic function defined in Eq.~\eqref{holomorphic} and $a_{k} = \vartheta_1 [ (k+q_1)/b_2,\omega]$. For more details on the symmetry conditions leading to Eq.~\eqref{flatAband} we refer to SM~\cite{Suppmat}. In Eq.~\eqref{flatAband}, we set the $K$ and $K'$ points at $\pm \bq_1$ respectively. Thanks to the following property of the theta-function $\vartheta_1 (-z,\omega) = - \vartheta_1 (z,\omega)$, it is readily checked that the poles of $\eta_{\bk \pm  \bq_1} (z)$ at $z=0$ cancel each other in Eq.~\eqref{flatAband}, as a result of Eq.~\eqref{equal-spinor}, and the wavefunction is finite everywhere. We note that, the corresponding unnormalized Bloch function ${\bm u}_{k} (\br) = \bm\psi_{\bk}(\br) e^{- i \bk \cdot \br}$ is $k$-holomorphic and thus constitutes an ideal flat band~\cite{Ledwith_2020,Wang_2021}. In addition, momentum space boundary conditions~\ref{k_space_BC} give $\phi_{k,b_1}=4\pi k/b_2+2\pi b_1/b_2$ and $\phi_{k,b_2}=0$ leading to a Chern number $C_A=2$ which gives the total Chern number $C=C_A+C_B=+1$ characteristic of the triangular regions centered at the ABA sites of the real-space pattern in Fig.~\ref{fig:chern_mosaic}b. Remarkably, the Chern $2$ band of Eq.~\eqref{flatAband} describes a color-entangled wavefunction~\cite{wang2022origin} and, upon translation of a lattice vector $\br_0\to\br_0+\ba_1$, the $k$-space zeros of $|\psi_{\bk2}(\br)|$ get swapped, see Fig.~\ref{fig:nodes_swapping}. 
The  emergence of the Chern bands $+2$ and $-1$  can be intuitively understood as a direct consequence of the original three Dirac cones of each layer, similar to twisted monolayer-bilayer
graphene~\cite{Ledwith_Khalaf_2022,Jie_hierarchyflatband}.

\paragraph*{AAA-stacking and domain wall lines--} The local Hamiltonian describing the AAA points is obtained by setting ${\bm \phi} =0$ in Eq.~\eqref{H_trilayer_staircase}. It satisfies all symmetries~\cite{Christophe_2019}: $C_{2z}T$, $C_{2x}$, $C_{3z}$ and particle-hole symmetry $P$, protecting the Dirac cones at $K$, $K'$ and $\Gamma$. $C_{2z}T$ furthermore enforces~\cite{Christophe_2019} a fully connected spectrum as an odd number of Dirac cones cannot form isolated minibands~\cite{Ahn_2019,Cano_2021,Yuncheng2023}.
\begin{figure}
    \centering
    \includegraphics[width=0.45\textwidth]{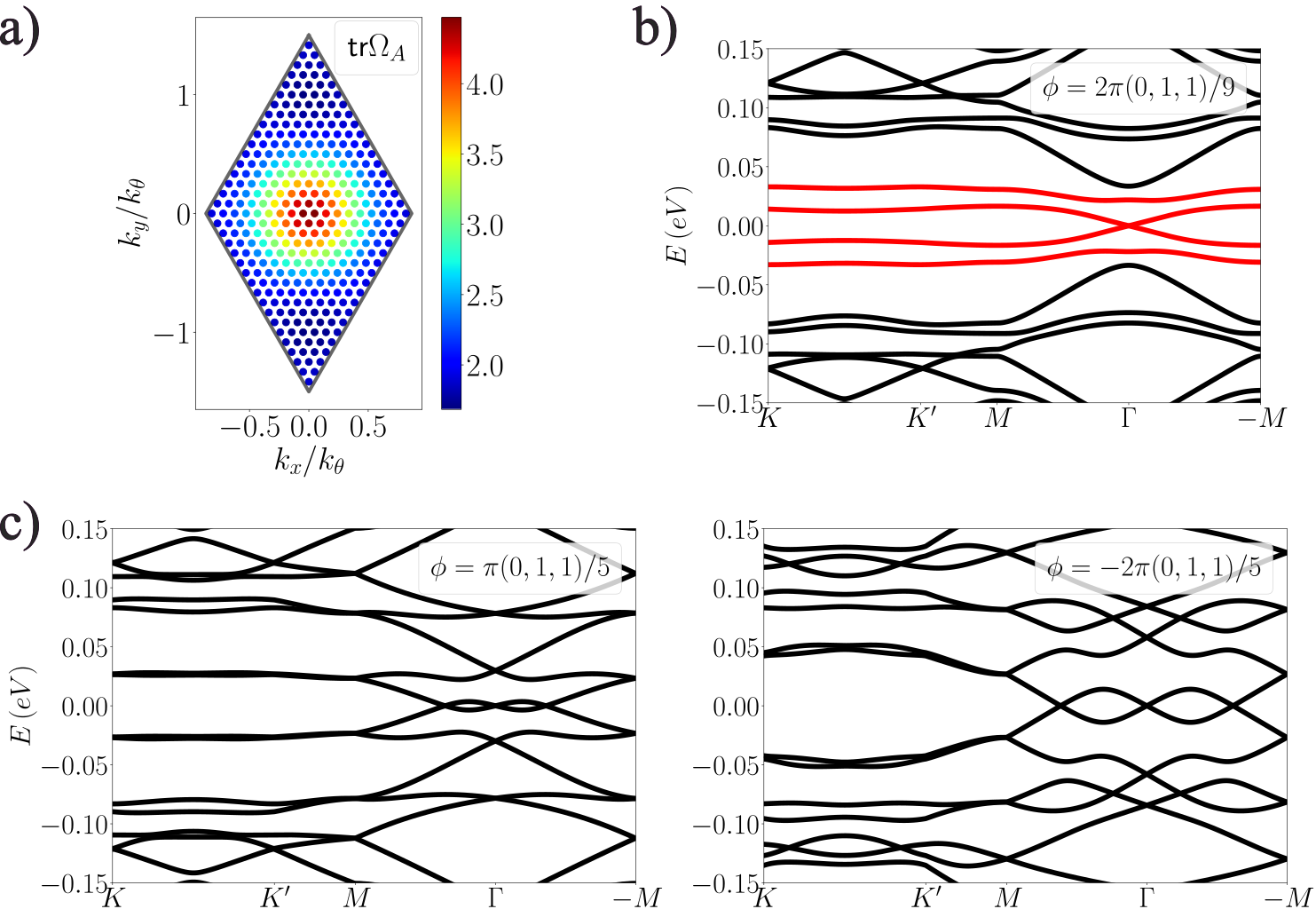}
    \caption{a) Trace of the non-Abelian Berry curvature for the A sublattice $\Lambda=+1$ (due to $C_{2z}T$ we have $\text{tr}\Omega_B=-\text{tr}\Omega_A$). We check numerically that the trace condition is satisfied. b) Spectrum obtained moving away from the AAA stacking along to the $\mathbf{a}^{\rm MM}_1-\mathbf{a}^{\rm MM}_2$ direction. A gap opens in the spectrum, we highlight in red the bands giving rise to a total Chern number $+1$. c) Bands obtained along the domain wall regions $\mathbf{a}^{\rm MM}_1+\mathbf{a}^{\rm MM}_2$ direction. The results are obtained at the magic angle $\theta_{AAA}\approx0.75^\circ$ and for $w_{\rm AA}=0.$} \label{fig:AAA_domains}
\end{figure}
In the chiral limit ($w_{\rm AA}=0$) the Hamiltonian $\mathcal H_{\rm AAA}$ in the basis $\Psi=\begin{pmatrix} \psi_1 & \psi_2 & \psi_3 & \chi_1 & \chi_2 & \chi_3 \end{pmatrix}^T$ takes the form:
  \begin{equation}
\label{H_AAA}
    \mathcal {H}_{\rm AAA}(\br)=\begin{pmatrix}
    0 & \mathcal D_2(\br)\\
   \mathcal D^\dagger_2(\br) & 0 
    \end{pmatrix},
\end{equation}
where the operators reads:
    \begin{equation}
\label{differential_operator_D_2}
    \mathcal D_2(\br)=\begin{pmatrix}
    -i\sqrt{2}\partial & U_{\omega^*}(\br) & 0 \\
    U_{\omega^*}(-\br) & -i\sqrt{2}\partial &U_{\omega^*}(\br)\\
   0 & U_{\omega^*}(-\br) & -i\sqrt{2}\partial
    \end{pmatrix},
\end{equation}
and $U_{\omega^*}(\br)=U^*_\omega(-\br)$. At the magic angle taking place at $\theta_{\rm AAA}\approx 0.75^\circ$, see red line in Fig.~\ref{fig:magic_flat_bands}b, the spectrum shown in Fig.~\ref{fig:magic_flat_bands}d is composed by a fourfold degenerate zero mode subspace and a renormalized Dirac cone located at $\Gamma$. Interestingly, the wavefunction of the zero modes can be exactly expressed in terms of meromorphic functions as shown in Ref.~\cite{popov2023magic}. We here focus on the topological properties of the fourfold degenerate flat band sector. 
Away from the $\Gamma$ point the flat bands are isolated and the degeneracy can be partially resolved by $\Lambda_z$ which gives two dimensional subspaces with opposite sublattice polarization $\Lambda_z=\pm1$. For a given sublattice the topological properties are characterized by the non-Abelian quantum geometric tensor $Q^{ab}_{nm}(\bk)=\braket{D_a u_{n\bk}}{D_b u_{m\bk}}$ with $D_a$ the covariant derivative~\cite{resta2020geometry}. The non-Abelian trace condition~\cite{parker2021fieldtuned} reads as $\Tr\text{tr} \,g = \text{tr}\,\Omega$ with $\Tr$ trace over space directions and $\text{tr}$ over the 2D subspace. Fig.~\ref{fig:AAA_domains}a shows the Berry curvature $\text{tr}\,\Omega_A$, we check numerically that the trace condition is satisfied everywhere at the exclusion of the $\Gamma$ point where the Berry curvature is ill-defined.  $C_{2z}T$ imposes that the two sublattice sectors yield opposite Berry curvature $\text{tr}\Omega_A=-\text{tr}\Omega_B$. The points AAA are however singular. The fourfold degeneracy is lifted by any small but finite ${\bm \phi}$ and the flat band sector with Chern number $\pm1$ is recovered, see Fig.~\ref{fig:AAA_domains}b. The Chern mosaic of Fig.~\ref{fig:chern_mosaic}b is thus largely governed by the topology of the ABA and BAB points.


Finally, the different topological regions extended around ABA and BAB meet along lines where the gaps to the remote bands vanish. These lines form a triangular lattice originating from the AAA lattice sites as shown in Fig.~\ref{fig:chern_mosaic}b. One can prove that, along these lines, the $C_{2x}$ symmetry, combined with $C_{3z}$ yields a fully connected band structure for Eq.~\eqref{H_trilayer_staircase}, as seen in Fig.~\ref{fig:AAA_domains}c. The proof, which follows Ref.~\cite{Christophe_2019}, is given in the SM~\cite{Suppmat}. However, breaking these symmetries can move these domain walls but not suppress them since the distinct topological domain must be separated by gap-closing contours.

\begin{figure}
    \centering
    \includegraphics[width=0.48\textwidth]{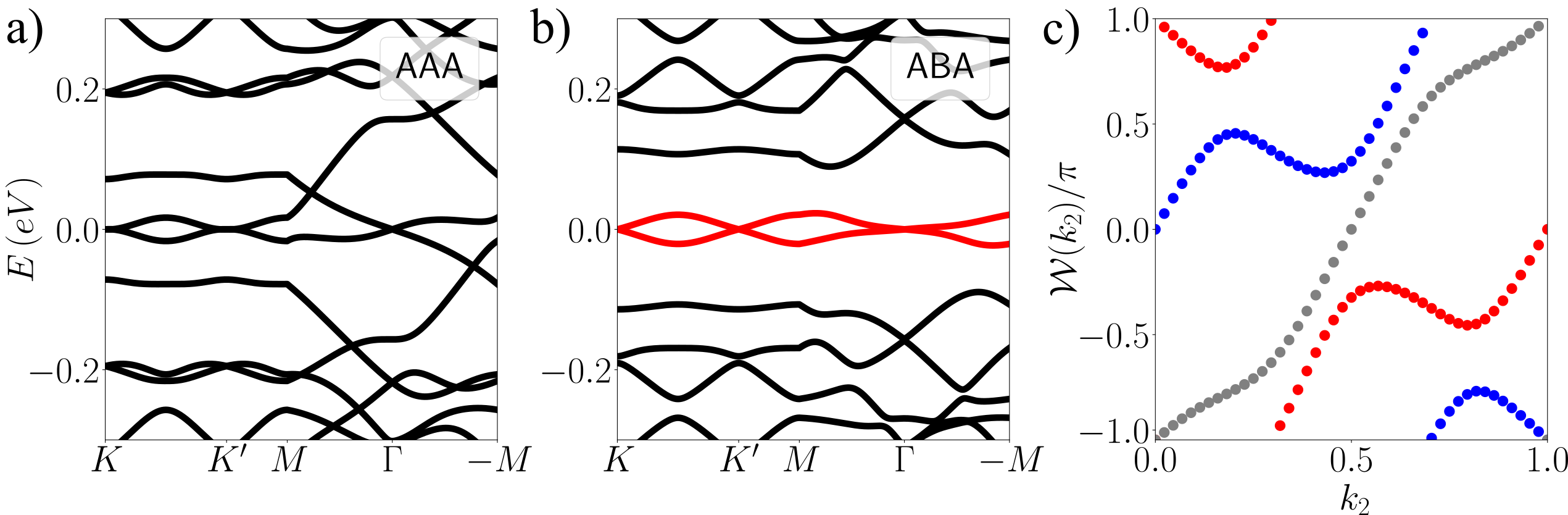}
    \caption{Results away from chiral limit ($w_{\rm AA}/w_{\rm AB}=0.7$) for twist angle $\theta=1.69^\circ$ corresponding to the magic angle for ABA stacking in the chiral limit. a) Fully connected spectrum for AAA stacking. b) Spectrum for ABA stacking, the flat bands separated from remote ones are highlighted in red. c) Wilson loop $\mathcal W(k_2)$ spectrum for the flat bands as a function of $k_2=\bk\cdot\ba_2/2\pi$. Red and blue dots shows the eigenvalues for the two flat bands, grey dots, instead, show the evolution of the center of mass characterized by a 2$\pi$ winding corresponding to a total Chern number $1$.} \label{fig:finite_wAA}
\end{figure}

\paragraph*{Stability away from the chiral limit ---} Our predictions formally derived in the chiral limit are stable and persists for finite value of $w_{\rm AA}$. We see that at finite $w_{\rm AA}$ the low-energy bands at ABA regions acquire a finite dispersion as shown in Fig.~\ref{fig:finite_wAA}b. 
However, the two flat bands highlighted in red in Fig.~\ref{fig:finite_wAA}b are characterized by a total Chern number $1$ as shown by the winding $2\pi$ of the center of mass of the Wilson loop $\mathcal W(k_2)$~\cite{Alexandradinata_2016}, gray dots in Fig.~\ref{fig:finite_wAA}c while red and blue dots show the evolution of the two eigenvalues. The Chern number mosaic pattern over the entire supermoiré space was previously presented in Fig.~\ref{fig:chern_mosaic}b with $w_{\rm AA}=0.8 w_{\rm AB}$.

\paragraph*{Conclusions ---} In summary, we showed that in equal-twist angle trilayer graphene separation between length scale leads to an interesting concept of supermoir\'e lattice where the local registry corresponds to twisting around AAA or ABA stacking but has a local-to-local variation over long range. 
The local adiabatic Hamiltonian which depends parametrically on the supermoir\'e lattice coordinate shows topologically distinct regions where the low-energy flat bands have finite and opposite Chern number. Remarkably, the non-vanishing Chern number at ABA regions originates from a zero modes composed by a Chern $+2$ color-entangled wavefunction and a Chern $-1$ Landau level like state. The intermediate domain wall regions originating from the AAA stacking configurations are characterized by a fully connected spectrum which is gapless at all energies. We conjecture that similar properties can be realized also for unequal twist angle configurations once the decoupling between slow and fast length scales is performed. The large energy gap between flat and remote bands compared to the typical Coulomb energy scale makes ABA stacking eTTG an ideal playground for studying Fractional Chern insulators in higher Chern number bands.   

\paragraph*{Acknowledgments ---} We are grateful to Jie Wang, Andrei Bernevig, Jed Pixley, Nicolas Regnault and Raquel Queiroz for insightful discussions. D.G. also acknowledge discussions held at the 2023 Quantum Geometry Working Group meeting that took place at the Flatiron institute where he was introduced to some of the concepts presented here. We acknowledge support by the French National Research Agency (project TWISTGRAPH, ANR-21-CE47-0018).
The Flatiron Institute is a division of the Simons Foundation.

\textit{Note added:} after the completion of the draft we become aware of the work by Trithep Devakul et al.~\cite{devakul2023magicangle} which overlapped with part of our results.

\bibliography{ref.bib}


\onecolumngrid
\newpage
\makeatletter 

\begin{center}
\textbf{\large Supplementary material for: `` \@title ''} \\[10pt]
Daniele Guerci$^1$, Yuncheng Mao$^2$, and Christophe Mora$^2$   \\
\textit{$^1$ Center for Computational Quantum Physics, Flatiron Institute, 162 5th Avenue, NY 10010, USA}\\
\textit{$^2$ Universit\'e Paris Cit\'e, CNRS,  Laboratoire  Mat\'eriaux  et  Ph\'enom\`enes  Quantiques, 75013  Paris,  France}
\end{center}
\vspace{20pt}

\setcounter{figure}{0}
\setcounter{section}{0}
\setcounter{equation}{0}

\renewcommand{\thefigure}{S\@arabic\c@figure}
\makeatother

\appendix


These supplementary materials contain the details of analytic calculations as well as additional numerical details supporting the results presented in the main text. It is structured as follows.  In Sec.~\ref{app:length_scale_decoupling} we introduce the local Hamiltonian providing the effective lattice model describing adiabatically the slowly varying supermoir\'e  length scale.
In Sec.~\ref{app:convention} we set the notation used in the main text. 
In Sec.~\ref{app:symetries_ABA} and~\ref{app:symetries_AAA} we provide the symmetries for the ABA and AAA stacking regions and in Sec.~\ref{app:Chern_2_flat} the symmetry constraint on the $C=2$ zero mode solution. Moreover, in Sec.~\ref{app:fully_domain_wall} we prove that the spectrum is fully connected along the high-symmetry lines connected AAA sites.
Sec.~\ref{app:kin} introduce complex notation useful for finding the zero mode solutions in the chiral limit. In Sec.~\ref{app:Jacobi} we provide some basic properties of the Jacobi functions. Finally, in Sec.~\ref{app:triple_product} we introduce the Wronskian which plays a crucial role for finding the analytical expression of the Chern $2$ flat band.   

\section{Length scales decoupling and local Hamiltonian}
\label{app:length_scale_decoupling}

In this section we recall some results derived in Ref.~\cite{Yuncheng2023} for generic relative twist $\theta_{12}$ and $\theta_{23}$. Specializing the discussion to the case of equal twist angle $\theta_1/\theta_2\simeq1$ we define the moir\'e modulation:
\begin{equation}
\bq_1= \bar{\theta} \, {\bf u}_z \wedge {\bf K}    
\end{equation}
 with ${\bf u}_z$ axis perpendicular to the TBG plane, ${\bf K} $ graphene Dirac point of length $|{\bf K}|=K_D=4\pi/3a_{\rm G}$, $\bar{\theta}=(\theta_1+\theta_2)/2$. The supermoir\'e wave vector is 
\begin{equation}
\delta {\bq}_1 = \bar{\theta}^2 \,  {\bf u}_z \wedge {\bf u}_z \wedge {\bf K}+\delta \theta \, {\bf u}_z \wedge {\bf K},
\end{equation}
where $\delta\theta=\theta_2-\theta_1$
is the angle deviation.  Their principal axis are rotated by $90^\circ$ with respect to each other in the equal-angle case $\delta \theta=0$ and parallel in the opposite limiting case of $\bar{\theta}^2 \ll \delta \theta$~\cite{PhysRevLett.125.116404}. It is worth stressing that the supermoir\'e pattern develops even when the twist angles are exactly equal $\theta_1 = \theta_2$. The single-particle band spectrum is described within a continuum model where the Dirac cones in each layer are coupled by the transverse tunneling of electrons. In the ${\bf K}$ valley, it reads:
\begin{equation}
\label{H_trilayer_staircase_sup}
    H_K =\begin{pmatrix}
    \hat{\bk}\cdot\bm \sigma & \alpha \sum_{j=1}^{3}T_j\,e^{-i\bq_j\cdot\br}e^{-i\delta\bq_j\cdot\br/2} & 0\\
    h.c. & \hat{\bk}\cdot\bm \sigma & \alpha \sum_{j=1}^{3}T_j\,e^{-i\bq_j\cdot\br}e^{+i\delta\bq_j\cdot\br/2} \\
    0 & h.c. & \hat{\bk}\cdot\bm \sigma
    \end{pmatrix}.
\end{equation}
Given $|\delta\bq_j|\ll|\bq_j|$ the slow variables are the phases: 
\begin{equation}
    \phi_j(\br)=\delta\bq_j\cdot\br/2
\end{equation}
which vary over the supermoir\'e lengthscale. On shorter scales, we can approximate them to be constant $\phi_j(\br)\simeq\phi_j$ introducing the local Hamiltonian: 
\begin{equation}
    H_{\rm eTTG}(\br,{\bm \phi}) =\begin{pmatrix}
    v_F\hat{\bk}\cdot\bm \sigma & T(\br,{\bm \phi}) & 0\\
    h.c. & v_F\hat{\bk}\cdot\bm \sigma & T(\br,-{\bm \phi}) \\
    0 & h.c. & v_F\hat{\bk}\cdot\bm \sigma
    \end{pmatrix},
\end{equation}
given in the manuscript. In fact we emphasize that $\phi_j$ are approximately constant on the moir\'e scale but still depend on the position ${\bf R}$ evolving on the supermoir\'e superlattice and leading to the Chern mosaic discussed in the manuscript.  

\section{Conventions}
\label{app:convention}

We defined $ \bq_j = ie^{2(j-1)\pi/3}$ where momenta are measured in unit of $k_\theta=\theta K_D$ with $K_D=4\pi/3a_{\rm G}$ and we employed the complex number notation. 
In unit of $k_\theta$ the reciprocal lattice vectors are given by: 
\begin{equation}
    \bm b_1=\bq_1-\bq_2=\sqrt{3}e^{i\pi/3},\quad \bm b_2=\bq_1-\bq_3=\sqrt{3}e^{2i\pi/3}.
\end{equation}
In unit of $1/k_\theta$ the lattice vectors are given by:
\begin{equation}
    \bm a_1=4\pi e^{i\pi/6}/3,\quad \bm a_2=4\pi e^{5i\pi/6}/3.
\end{equation}
We define $K=\bq_1$ and $K'=- \bq_1$.
\begin{equation}
\label{conventions}
\begin{array}{c||c|c|c|c|c|c} 
r & a_1 &  a_2 &  a_1+a_2 &  r_0 & r_1 &  r_2 \\\hline\hline
e^{i K\cdot r} & \omega & \omega & \omega & 1 & \omega & \omega^* \\\hline
e^{iK'\cdot r } & \omega^* & \omega^* & \omega^* & 1 & \omega^* &  \omega 
\end{array} \, \, , \quad \omega = e^{2i\pi/3} ,
\end{equation}
where $r_j=C^j_{3z}r_0$ with $j=0,1,2$ and $r_0=(a_1-a_2)/3$. Finally, the original Hamiltonian is written in the basis: 
\begin{equation}
\label{original_basis}
    \Phi=\begin{pmatrix}
     \psi_1 &
     \chi_1 &
     \psi_2 &
     \chi_2 &
     \psi_3 &
     \chi_3
    \end{pmatrix}^T,
\end{equation}
where $1,2,3$ refer to top, middle and bottom layers, respectively. The two components $\psi$ and $\chi$ refer to A and B sublattice, respectively.

\section{Symmetries and protected Dirac cones for ABA stacking}
\label{app:symetries_ABA}

In the following we discuss the symmetry properties of eTTG with respect to the ABA stacking configuration. 
The Hamiltonian representative of the ABA region is obtained setting the values of the phases ${\bm\phi}_{\rm ABA}=(\phi_1,\phi_2,\phi_3)= (0,\varphi,-\varphi)$ with $\varphi=2\pi/3$, notice that one of the phases say $\phi_1$  can be gauged away by global phase shift $\phi_j\to\phi_j+\phi_0$. 
The Hamiltonian in the basis~\ref{original_basis} reads:
\begin{equation}
\label{H_ABA_supp}
    \mathcal {H}_{\rm ABA} \equiv H_{\rm eTTG}({\bm \phi}_{\rm ABA}) =\begin{pmatrix}
    \hat{\bk}\cdot\bm \sigma  & T_{\omega^*}(\br) & 0\\
    h.c. & \hat{\bk}\cdot\bm \sigma & T_\omega(\br) \\
    0 & h.c. & \hat{\bk}\cdot\bm \sigma 
    \end{pmatrix}.
\end{equation}
where the tunneling matrices are:
\begin{equation}
T_{\omega^*}(\br)\equiv T(\br,{\bm\phi})=\alpha\sum^3_{j=1}\left(\omega^*\right)^{j-1}T_je^{-i\bar \bq_j\cdot\br},\quad T_{\omega}(\br)\equiv T(\br,-{\bm\phi})=\alpha\sum^3_{j=1}\omega^{j-1}T_je^{-i\bar \bq_j\cdot\br}.
\end{equation}
The $C_{3z}$ symmetry takes the form: 
\begin{equation}
    C_{3z}=\begin{pmatrix}
    \omega^*e^{i\varphi\sigma_z} & 0 & 0 \\
    0 & e^{i\varphi\sigma_z} & 0 \\
    0 & 0 & \omega^*e^{i\varphi\sigma_z} 
    \end{pmatrix}= \begin{pmatrix}
    \omega^* & 0 & 0 \\
    0 & 1 & 0 \\
    0 & 0 & \omega^*
    \end{pmatrix}\otimes e^{i\varphi\sigma_z}=\Omega\otimes e^{i\varphi\sigma_z},
\end{equation}
and 
\begin{equation}
     \mathcal {H}_{\rm ABA}(C_{3z}\br)=C_{3z}
     \mathcal {H}_{\rm ABA}(\br)C^\dagger_{3z}.
\end{equation}
We observe that $C_{2x}$ is defined as: 
\begin{equation}
    C_{2x}=\begin{pmatrix}
        0 & 0 & 1 \\
        0 & 1 & 0 \\
        1 & 0 & 0
    \end{pmatrix}\otimes \sigma_x=\mathcal{M}_x\otimes\sigma_x,
\end{equation}
is broken, since under $C_{2x}$ we map the ABA to the BAB stacking. This can be readily realized observing that: 
\begin{equation}
C_{2x}{\mathcal H_{\rm ABA}}(\br)C^\dagger_{2x}=1\otimes(C_{2x}\bk)\cdot\bm \sigma+\begin{pmatrix}
    0 & T_{\omega}(C_{2x}\br) & 0\\
    h.c. & 0 & T_{\omega^*}(C_{2x}\br) \\
    0 & h.c. & 0 
    \end{pmatrix}= {\mathcal H_{\rm BAB}}(C_{2x}\br).
\end{equation}
Similarly, we find: 
\begin{equation}
    C_{2z}T{\mathcal H_{\rm ABA}}(\br)(C_{2z}T)^\dagger=-1\otimes\bk\cdot\bm \sigma+\begin{pmatrix}
    0 & T_{\omega}(-\br) & 0\\
    h.c. & 0 & T_{\omega^*}(-\br) \\
    0 & h.c. & 0 
    \end{pmatrix}={\mathcal H_{\rm BAB}}(-\br).
\end{equation}
We now notice that the combination $C_{2x}C_{2z}T=C_{2y}T$
\begin{equation}
    C_{2y}T =\mathcal M_x\otimes\sigma_0 \mathcal K
\end{equation}
is a symmetry of the model:
\begin{equation}
C_{2y}T{\mathcal H_{\rm ABA}}(\br)(C_{2y}T)^\dagger=1\otimes\left(C_{2y}\bk\right)\cdot\bm \sigma+\begin{pmatrix}
    0 & T_{\omega^*}(C_{2y}\br) & 0\\
    h.c. & 0 & T_{\omega}(C_{2y}\br) \\
    0 & h.c. & 0 
    \end{pmatrix}= {\mathcal H_{\rm ABA}}(C_{2y}\br).
\end{equation}
We summarize the symmetries in Table~\ref{tab:Symmetries_AAA_ABA}.

\begin{table}[]
\centering
\begin{tabular}{|c||c|c||c|c|c|c|}
\hline
ABA& Original &Chiral& $  k_+$&$  k_-$&$r_+$&$r_-$  \\
\hline\hline
$C_{3z}$& $\Omega\otimes e^{i\varphi\sigma_z}$ & $e^{i\varphi\sigma_z}\otimes\Omega$ & $\omega$&$\omega^*$&$\omega$&$\omega^*$ \\
\hline
$P$& $\mathcal{M}\otimes\sigma_0$ & $\sigma_0\otimes\mathcal{M}$ &$-$&$-$&$-$&$-$ \\
\hline
$C_{2x}C_{2z}T$& $\mathcal{M}\otimes\mathcal K$ & $\mathcal K\otimes\mathcal{M}$ &$-k_-$&$-k_+$&$-r_-$&$-r_+$ \\
\hline\hline
$\Lambda_z$&$\tau_0\otimes\sigma_z$ & $\sigma_z\otimes\tau_0$ &$+$&$+$&$+$&$+$ \\
\hline
$P \otimes \Lambda_z$ & $\mathcal{M}\otimes\sigma_z$ & $\sigma_z\otimes\mathcal M$ & $-$ & $-$ & $-$ & $-$ \\
\hline
\hline 
\hline
AAA& Original &Chiral & $k_+$&$  k_-$&$r_+$&$r_-$ \\
\hline\hline
$C_{2x}$& $\mathcal{M}_x\otimes\sigma_x$ & $\sigma_x\otimes\mathcal{M}_x$ & $k_-$&$k_+$&$r_-$&$r_+$ \\
\hline
$C_{3z}$& $\mathbf 1\otimes e^{i\varphi\sigma_z}$ &  $ e^{i\varphi\sigma_z}\otimes\mathbf 1$  & $\omega$&$\omega^*$&$\omega$&$\omega^*$ \\
\hline
$C_{2z}T $& $\mathbf 1\otimes \sigma_x \mathcal K$ &  $ \sigma_x\otimes\mathbf 1 \mathcal K$  & $+$&$+$&$-$&$-$ \\
\hline
$P$&   $\mathcal{M}\otimes\sigma_0 $& $\sigma_0\otimes\mathcal{M}$ 
 &$-$&$-$&$-$&$-$ \\
\hline\hline
$\Lambda_z$&$\tau_0\otimes\sigma_z$ & $\sigma_z\otimes\tau_0$ &$+$&$+$&$+$&$+$ \\
\hline
$P \otimes \Lambda_z$ & $\mathcal{M}\otimes\sigma_z$ & $\sigma_z\otimes\mathcal M$ & $-$ & $-$ & $-$ & $-$ \\
\hline
\end{tabular}
\caption{\textbf{Symmetries of the local Hamiltonian for the two different high-symmetry stacking configurations for the K-valley}. Original and chiral shows the matrix expression of the symmetry in the two different basis. Finally, last columns show how the symmetry acts on $k_{\pm}=k_x\pm ik_y$ and $r_\pm=x\pm iy$. We further notice that $\mathcal K$ is the complex conjugation operator and the ABA stacking breaks $C_{2z}T$ and $C_{2x}$ symmetries. The combination $C_{2x}C_{2z}T$ is still a symmetry of the model.
} 
\label{tab:Symmetries_AAA_ABA}
\end{table}

To the aim of showing the protection of three Dirac cones at $\Gamma$, $K$ and $K'$ we further discuss the spatial symmetries in the ABA configuration and the symmetry group they generate. As detailed above, the model exhibits two spatial symmetries in that case, $C_{3z}$ and $C_{2y} T$. They form the magnetic space group $P32'1$ ($\#150.27$ in the BNS notation). The irreducible representations at the high-symmetry momenta $\Gamma$ and $K$ ($K'$) verify the $C_3$ point group character table and are all one-dimensional. The absence of two-dimensional representations noticeably prevents the stabilization of Dirac cones in the spectrum.

It can be understood more directly by considering an eigenstate of $C_{3z}$,  $C_{3z} | \psi_+ \rangle = \omega | \psi_+ \rangle$. From the relation
\begin{equation}
    (C_{2 y} T) C_{3 z} (C_{2 y} T) = C_{3 z}^{-1}, 
\end{equation}
we obtain $C_{3 z} C_{2 y} T | \psi_+ \rangle = \omega \, C_{2 y} T | \psi_+ \rangle$. $C_{2 y} T$ thus does not circulate between the eigenstates of $C_{3z}$ and cannot protect a twofold degeneracy.

The three zero-energy Dirac cones arising in the band spectrum of ABA trilayer graphene are therefore only stable in the presence of the additional particle-hole symmetry $P$. Since $P$  and $C_{3 z}$ commute, applying $P$ on a given state does not change its $C_{3z}$ eigenvalue. Therefore, if the spectrum, at $\Gamma$ or $K/K'$, hosts two states with $C_{3z}$ eigenvalues $\omega$, $\omega^*$ near charge neutrality (and no other states), particle-hole symmetry $P$ automatically pins these two states at zero energy. This is proven by contradiction: if we assume that the two states sit at opposite non-vanishing energies, then $P$ permutes them. This is however impossible since $P$ cannot change the $C_{3z}$ eigenvalue which completes the proof. In fact, $P$ restricted to these two states must be the identity as it commutes with $C_{3 z}$. It further shows that breaking $C_{3 z}$ does not lift the Dirac crossings as the trivial (identity) representation of $P = \mathbb{I}_2$ cannot deform continuously to the traceless $\sigma_x$ matrix permuting states with opposite non-zero energies.

In summary, the Dirac cone at $\Gamma$ is explained using $C_{3 z}$ and $P$. $K$ and $K'$ are however not stable under $P$ - which permutes $K$ and $K'$ - and the stability of their Dirac cones must come from a different operator. Since the symmetry $C_{2 y} T$ also permutes $K$ and $K'$, the combination $P' = P C_{2 y} T$ leaves them invariant and acts as a (anti-unitary) particle-hole operator. Using that 
\begin{equation}
    P' C_{3 z} P' = C_{3 z}^{-1},
\end{equation}
one verifies that $P'$ cannot permute the $C_{3 z}$ eigenvalue between $\omega$ and $\omega'$ which implies that an isolated pair of states at $K$ or $K'$ with $C_{3 z}$ eigenvalues $(\omega,\omega^*)$ is degenerate and pinned at zero energy. In this subspace, $P' = \mathcal{K} \, \mathbb{I}_2$ and the Dirac cone survives a breaking of $C_{3 z}$.

\subsection{Symmetry constraints on the Chern $2$ flat band}
\label{app:Chern_2_flat}

Here we provide the symmetry constraint on the $C=2$ zero mode that leads to the analytical expression~\ref{flatAband} given in the main text.

To start with we observe that the $C_{3z}$ symmetry implies $\psi_{\Gamma 1/3}(\br)=\omega \psi_{\Gamma 1/3}(C_{3z}\br)$ and $\psi_{\Gamma 2}(\br)= \psi_{\Gamma 2}(C_{3z}\br)$ resulting in the condition $\psi_{\Gamma 1/3}(0)=0$. However, differently from the B-sublattice component we find $\psi_{\Gamma 2}(0)\neq0$ at the magic angle. To proceed further we now observe that ${\cal M} \mathcal D^\dagger_1(- \br){\cal M} = - \mathcal D^\dagger_1( \br)$ with $\mathcal M$ defined in Eq.~\ref{particle_hole_symmetry} of the manuscript. At the $\Gamma$ point, it leads to $\psi_{\Gamma2}(\br) = \psi_{\Gamma2}(-\br)$, and $\psi_{\Gamma1}(\br) = \psi_{\Gamma3}(-\br)$ which connects the top and bottom components at inversion-symmetric positions.
Looking at $K$  and $K'=-K$ points the $C_{3z}$ symmetry leads to the identities $\psi_{\pm K1/3}(\br)=\psi_{\pm K1/3}(C_{3z}\br)$ and $\psi_{\pm K2}(\br)= \omega^* \psi_{\pm K2}(C_{3z}\br)$ consistent with $\psi_{K} = (1,0,0)$ and $\psi_{K'} = (0,0,-1)$ at $w_{\rm AB}=0$. A direct consequence is that $\psi_{\pm K2}(0) = 0$. In contrast to $C_{3z}$, the particle-hole symmetry $\mathcal M$ couples $K$ and $K'$ we find $\psi_{K 1/3}(\br)= - \psi_{-K 1/3}(-\br)$ and $\psi_{K 2}(\br)=  \psi_{- K 2}(-\br)$ resulting in $\psi_{\pm K 1}(0) = - \psi_{\mp K3}(0)$. Moreover and remarkably, we  numerically find that, right at the magic angle $\theta_{\rm ABA}$
\begin{equation}
  \psi_{K 1}(0) = -  \psi_{K 3}(0),
\end{equation}
which results in the fact that the  two spinors $\psi_{K} (\br = 0 )$ and $\psi_{K'} (\br = 0)$ become in fact identical. 
We now generalize the argument in Ref.~\cite{popov2020hidden,Stern_2023} defining the Wronskian for  the case of three layers.
To this aim we consider three zero mode solutions ${\bm \psi}_{\bk_j}$ with momenta ${\bk}_{1,2,3}$ and we introduce the triple product 
\begin{equation}
\label{triple_product}
   W (\br) = {\bm \psi}_{\bk_1}(\br) \cdot \left [{\bm \psi}_{\bk_2}(\br) \times {\bm \psi}_{\bk_3}(\br) \right],  
\end{equation}
which satisfies $\bar \partial W(\br) = 0$, see Sec.~\ref{app:triple_product}, proving the holomorphy of $W(\br)$. Since $W(\br)$ also satisfies Bloch periodicity (being the product of three Bloch-periodic functions) and cannot have any pole in the complex plane, by Lioville's theorem it must be constant in space. Taking $\bk_{2} = K$ and $\bk_{3} = K'$ and an arbitrary momentum $\bk_{1}$, the quantity $W (\br=0)=0$ since ${\bm \psi}_{K} (0 )$ and ${\bm \psi}_{K'} (0)$ are collinear as discussed above. Therefore the triple product $W (\br)$ vanishes in this case and the three spinors ${\bm \psi}_{\bk_1}(\br)$, ${\bm \psi}_{K}(\br)$ and ${\bm \psi}_{K'}(\br)$ are coplanar for all $\br$. Since this is true for arbitrary $\bk_1$, it implies that all zero-energy states belong to the same plane generated by ${\bm \psi}_{K}(\br)$ and ${\bm \psi}_{K'}(\br)$ and the triple product $W (\br)$ is identically vanishing irrespective of the values of $\br$ and $\bk_{1,2,3}$.
Thus, an analytical expression also emerges from the previous analysis and the wavefunction for the  A-polarized flat band takes the form (up to $\bk$-dependent prefactor) in Eq.~\ref{flatAband} of the manuscript.

\section{Symmetries for AAA stacking}
\label{app:symetries_AAA}

In the following we discuss the symmetry properties of eTTG with respect to the AAA stacking configuration. 

The AAA stacking configuration ${\bm\phi}_{\rm AAA}=(0,0)$ recently discussed in the preprint~\cite{popov2023magic} is described by the Hamiltonian~\cite{Christophe_2019}: 
\begin{equation}
\label{H_trilayer_staircase_ABA}
    \mathcal {H}_{\rm AAA} \equiv H_{\rm eTTG}({\bm \phi}_{\rm AAA}) =\begin{pmatrix}
    \hat{\bk}\cdot\bm \sigma  & T_0(\br) & 0\\
    h.c. & \hat{\bk}\cdot\bm \sigma & T_0(\br) \\
    0 & h.c. & \hat{\bk}\cdot\bm \sigma 
    \end{pmatrix}.
\end{equation}
In this case the tunneling matrix simply reads: 
\begin{equation}
    T_0(\br)=\alpha\sum^3_{j=1}T_je^{-i\bar \bq_j\cdot\br},
\end{equation}
under $C_{3z}$ it transforms as: 
\begin{equation}
    e^{-i\varphi\sigma_z}T_0(C_{3z}\br)e^{i\varphi\sigma_z}=T_0(\br ).
\end{equation}
We readily understand that the three layers in this case are characterized by the same $C_{3z}$ eigenvalue. Thus, the $C_{3z}$ operator reads: 
\begin{equation}
    C_{3z}=\begin{pmatrix}
    e^{i\varphi\sigma_z} & 0 & 0 \\
    0 & e^{i\varphi\sigma_z} & 0 \\
    0 & 0 & e^{i\varphi\sigma_z} 
    \end{pmatrix}=\mathbf 1\otimes e^{i\varphi\sigma_z} ,
\end{equation}
and 
\begin{equation}
     \mathcal {H}_{\rm AAA}(R_\varphi\br)=C_{3z}
     \mathcal {H}_{\rm AAA}(\br)C^\dagger_{3z}.
\end{equation}
For the sake of completeness we list the remaining symmetries: 
\begin{equation}
    C_{2x}=\begin{pmatrix}
        0 & 0 & 1 \\
        0 & 1 & 0 \\
        1 & 0 & 0
    \end{pmatrix}\otimes \sigma_x=\mathcal{M}_x\otimes\sigma_x,
\end{equation}
so that 
\begin{equation}
    \mathcal H_{\rm AAA}(C_{2x}\br)=C_{2x}\mathcal H_{\rm AAA}(\br)C^\dagger_{2x}.
\end{equation}
In the previous expression $\mathcal M_x$ corresponds to the operator which flips bottom and top layer leaving the mid one invariant. Table~\ref{tab:Symmetries_AAA_ABA} summarizes the symmetries of the model. We emphasize that the symmetries $P$, $C_{3z}$ and $C_{2z}T$ protect the Dirac cones at $\Gamma$, $K$ and $K'$ of the moir\'e Brillouin zone.

\subsection{Fully connected spectrum at the domain wall regions}
\label{app:fully_domain_wall}

In this section we prove by symmetry arguments that the spectrum is fully connected at the domain wall regions. 
To this aim we shift the position of the middle layer along the $x$ axis with respect to the AAA stacking. By construction, this choice, with $\phi_2 = \phi_3$ where we used a gauge transformation to set $\phi_1=0$, does not break the $C_{2 x}$ symmetry but breaks $C_{3 z}$ and $C_{2 z} T$. The particle-hole symmetry $P$~\ref{particle_hole_symmetry} is also not broken such that there is particle-hole symmetry at all exceptional points. The corresponding moire spectrum is displayed in the right panel of Fig.~\ref{fig:AAA_domains}c in the manuscript for $\phi_2 = \phi_3 =- 2 \pi /5$ where a gap opening at $K$ ($K'$) is observed resulting from the breaking of $C_{3 z}$ and $C_{2 z} T$. In contrast, $P$ still applies which stabilizes the Dirac point at $\Gamma$.

\begin{table}
\begin{center}
\begin{tabular}{ |c|c|c|c||c|c|c| } 
 \hline
  & $E$ \quad&\quad 2$C_3$ \quad&\quad 3$C_2$ \quad &\quad &\quad $E$ &\quad 3$C_2$ \\ 
 \hline\hline
 $\quad \Gamma_1\quad$  &\quad  1 \quad &\quad  1\quad  &\quad  1\quad & \quad $M_1$ &\quad  1 &\quad 1  \\ 
 \hline
 $\quad \Gamma_2\quad $ & \quad 1 \quad &\quad  1\quad  &\quad  -1\quad & \quad $M_2$ &\quad 1 &\quad -1  \\
 \hline
 $\quad \Gamma_3\quad $ & \quad 2\quad  & \quad -1\quad  & \quad 0\quad & \quad &\quad &\quad   \\
 \hline
\end{tabular}
\caption {Character table of $C_{3v}$ at $\Gamma$ and $M$. $E$, $C_3$ and $C_2$ represent the conjugation classes of the identity, $C_{3z}$ and $C_{2x}$, respectively.} \label{app:character_table} 
\end{center}
\end{table}

Furthermore, we also see that the bands are all connected. The proof of this full connectivity goes along the same lines as in Ref.~\cite{Christophe_2019}. Suppose we take an isolated set of $N$ bands - $P$ allows us to choose this set symmetrically around zero energy. At the $\Gamma$ point, there are two one-dimensional representations of $C_{2x}$, $\Gamma_1$ and $\Gamma_2$, with the characters $+1$ and $-1$ see Tab.~\ref{app:character_table}. $P$ implies a third two-dimensional representation (the Dirac point) with character $0$, analytical continued from $C_{2 x} = \sigma_x$ as obtained for $\phi_1=\phi_2=0$. At the $M$ point and on the $\Gamma$-$M$ line, again two representations $M_1=+1$, $M_2=-1$ see Tab.~\ref{app:character_table}. Writting that $m_{\Gamma_1} + m_{\Gamma_2} + 2 m_{\Gamma_3} = N$ (where $m_{\Gamma}$ is the multiplicity of the representation $\Gamma$), $m_{M_1} + m_{M_2} = N$, and that the full character is preserved along the $\Gamma$-$M$ line, namely
\begin{equation}
  m_{\Gamma_1} - m_{\Gamma_2} = m_{M_1} - m_{M_2},
\end{equation}
we find
\begin{equation}\label{character-id}
  m_{M_2} = m_{\Gamma_2} + m_{\Gamma_3}.
\end{equation}
This last expression can be checked modulo two on the zero-energy axis - as particle-hole partners within the $N$ bands share the same representation and thus count as a multiple of two, {\it i.e.} do not change the parity. The proof by contradiction is completed using that, for $w_{\rm AA}=w_{\rm AB}=0$, there is a Dirac point at $\Gamma$ and not at $M$ such that $m_{\Gamma_3} = 1$ but $m_{M_2} = m_{\Gamma_2}=0$, in contradiction with Eq.~\eqref{character-id}.


\section{The kinetic term in complex notation and chiral limit tunneling matrices}
\label{app:kin}

The kinetic term reads: 
\begin{equation}
        \hat{\bk}\cdot\bm \sigma=\begin{pmatrix}
            0& k_x-ik_y\\
            k_x+ik_y & 0 
        \end{pmatrix}, \quad k_a = - i \partial_a/k_\theta,
\end{equation}
where $k_\theta=\theta K_D$. We use the complex notation: 
\begin{equation}
    z=k_\theta\frac{x+iy}{\sqrt{2}},\quad \bar z=k_\theta\frac{x-iy}{\sqrt{2}},
\end{equation}
so that $x=(z+\bar z)/[\sqrt{2}k_\theta]$ and $y=-i(z-\bar z)/[\sqrt{2}k_\theta]$.
Finally, we have: 
\begin{equation}
    \partial_x-i\partial_y=\sqrt{2}k_\theta \partial ,\quad \partial_x+i\partial_y=\sqrt{2}k_\theta \bar \partial. 
\end{equation}
As a result we have 
\begin{equation}
        \hat{\bk}\cdot\bm \sigma=\begin{pmatrix}
            0& -i\sqrt{2}\partial\\
            -i\sqrt{2}\bar \partial & 0 
        \end{pmatrix}.
\end{equation}
In the chiral limit we have: 
\begin{equation}
\begin{split}
    T_1=\alpha\begin{pmatrix}
        0 & 1 \\
        1 & 0 
    \end{pmatrix},\,T_2=\alpha\begin{pmatrix}
        0 & \omega^* \\
        \omega & 0 
    \end{pmatrix},\,T_3=\alpha\begin{pmatrix}
        0 & \omega \\
        \omega^* & 0 
    \end{pmatrix},
\end{split}
\end{equation}
where $\alpha=w_{\rm AB}/v_Fk_\theta$.
As a result of the different $C_{3z}$ properties at the AAA and ABA-stacking we have three different scalar potentials: 
\begin{equation}
\begin{split}
   U_0(\br)=\alpha\sum^{3}_{j=1} e^{-i\bq_j\cdot\br}, \, U_\omega(\br)=\alpha\sum^{3}_{j=1} \omega^{j-1}e^{-i\bq_j\cdot\br},\,U_{\omega^*}(\br)=\alpha\sum^{3}_{j=1} \left(\omega^*\right)^{j-1}e^{-i\bq_j\cdot\br}.
\end{split}
\end{equation}
We have: 
\begin{equation}
 \begin{split}
    T_0=\begin{pmatrix}
        0 & U_{\omega^*} \\
        U_\omega & 0 
    \end{pmatrix},\, T_\omega= \begin{pmatrix}
        0 & U_{0} \\
        U_{\omega^*} & 0 
    \end{pmatrix},\, T_{\omega^*}= \begin{pmatrix}
        0 & U_{\omega} \\
        U_0 & 0 
    \end{pmatrix}.
 \end{split}   
\end{equation}
Moreover, we have the following relations: 
\begin{equation}
\begin{split}
    U^*_0(\br)=U_0(-\br),\,U^*_\omega(\br)=U_{\omega^*}(-\br),\,U^*_{\omega^*}(\br)=U_{\omega}(-\br).
 \end{split}   
\end{equation}

\section{Basic properties of Jacobi theta function}
\label{app:Jacobi}

In the maintext we have utilized the Jacobi theta-function:
\begin{equation}
    \vartheta_1( z , \omega) = \sum_{n \in \mathbb{Z} } e^{i \pi \omega (n+1/2)^2} e^{2 i \pi (z-1/2)(n+1/2)}.
\end{equation}
Here we recall some basic properties useful to determine the boundary conductions of the ideal flat bands.
We first notice that 
\begin{equation}
\begin{split}
    &\vartheta_1[z+\omega,\omega]=-e^{-i\pi\omega}e^{-2\pi i z}\vartheta_1[z,\omega],\\
    & \vartheta_1[z-\omega,\omega]=-e^{-i\pi\omega}e^{2\pi i z}\vartheta_1[z,\omega].
\end{split}
\end{equation}
In addition we also need: 
\begin{equation}
\vartheta_1[z\pm 1,\omega]=-\vartheta_1[z,\omega].
\end{equation}
Other useful relations are:
\begin{equation}
\begin{split}
    &\vartheta_1[\bar z-\omega^*,-\omega^*]=-e^{i\pi\omega^*}e^{-2\pi i \bar z}\vartheta_1[\bar z,-\omega^*],\\
    & \vartheta_1[\bar z+\omega^*,-\omega^*]=-e^{i\pi\omega^*}e^{2\pi i \bar z}\vartheta_1[\bar z,-\omega^*].
\end{split}
\end{equation}
Finally, we also have 
\begin{equation}
\vartheta_1[\bar z\pm 1,-\omega^*]=-\vartheta_1[\bar z,-\omega^* ].
\end{equation}

\section{Wronskian for the ABA-case: counting the number of independent components}
\label{app:triple_product}

In this Section we introduce the Wronskian~\cite{popov2020hidden,Stern_2023} which determines the vector space dimension of the zero mode solution in $\bk$-space. 
Given three solutions of the zero mode equations
\begin{equation}
    \mathcal D_1{\bm\chi}_{\bk}=0,\quad \mathcal D^\dagger_1{\bm\psi}_{\bk}=0,
\end{equation}
$\bm\psi_{a,b,c}$ and $\bm\chi_{a,b,c}$ with $a,b,c$ denoting different $\bk$ points we define the triple product: 
\begin{equation}
W=\bm\psi_a\cdot\left(\bm\psi_b\times\bm\psi_c\right),\quad \overline{ W}=\bm\chi_a\cdot\left(\bm\chi_b\times\bm\chi_c\right).
\end{equation}
Generically $W=W(z,\bar z)$ and $\overline{W}=\overline{W}(z,\bar z)$, in the following we show that the two quantities are holomorphic and antiholomorphic functions, respectively. To this aim we first observe that the zero mode equation can be written as: 
\begin{equation}
 \left(   \bar\partial +\overline{A}\right)\bm\psi_{\bk}=0,\quad \left(  \partial +A\right)\bm\chi_{\bk}=0,
\end{equation}
where we have introduced the gauge fields: 
\begin{equation}
    \overline{ A} =\frac{i}{\sqrt{2}}\begin{pmatrix}
    0 & U_0(\br) & 0 \\
    U_{\omega^*}(-\br) & 0 & U_{\omega^*}(\br)\\
   0 & U_0(-\br) & 0
    \end{pmatrix},\quad A=\frac{i}{\sqrt{2}}\begin{pmatrix}
    0 & U_\omega(\br) & 0 \\
    U_{0}(-\br) & 0 & U_{0}(\br)\\
   0 & U_\omega(-\br) & 0
    \end{pmatrix}.
\end{equation}
Without losing generality we now focus on $\overline{W}$, in this case: 
\begin{equation}
    \partial \overline{W}=\sum_{ljk}\varepsilon_{ljk}\partial\left[\chi_{a,l}\chi_{b,j}\chi_{c,k}\right]=-\sum_{ljks}\varepsilon_{ljk}A_{ks}\left[\chi_{b,l}\chi_{c,j}\chi_{a,s}-\chi_{a,l}\chi_{c,j}\chi_{b,s}+\chi_{a,l}\chi_{b,j}\chi_{c,s}\right].
\end{equation}
Now we observe that $A$ is off-diagonal. Thus, we have: 
\begin{equation}
\begin{split}
    \partial \overline{W}=&-\sum^{k\neq s}_{ks}A_{ks}\sum_j\left[ \varepsilon_{sjk}\left( \chi_{b,s}\chi_{c,j}\chi_{a,s}-\chi_{a,s}\chi_{c,j}\chi_{b,s}+\chi_{a,s}\chi_{b,j}\chi_{c,s}\right)+\right.\\
    &\left.\varepsilon_{jsk}\left(\chi_{b,j}\chi_{c,s}\chi_{a,s}-\chi_{a,j}\chi_{c,s}\chi_{b,s}+\chi_{a,j}\chi_{b,s}\chi_{c,s}\right)\right]\\
    =&-\sum^{k\neq s}_{ksj}A_{ks} \varepsilon_{sjk}\left[  \chi_{b,s}\chi_{c,j}\chi_{a,s}-\chi_{a,s}\chi_{c,j}\chi_{b,s}+\chi_{a,s}\chi_{b,j}\chi_{c,s} -\chi_{b,j}\chi_{c,s}\chi_{a,s}+\chi_{a,j}\chi_{c,s}\chi_{b,s}-\chi_{a,j}\chi_{b,s}\chi_{c,s}\right]=0.
\end{split}
\end{equation}
As a result $\overline{W}=\overline{W}(\bar z)$ is antiholomorphic function. Similarly, it is possible to prove $\bar \partial W=0$ so that $W=W(z)$. Since $W$ is regular it must be a constant. The only constant consistent with the boundary conditions is zero. The latter condition implies that the zero mode solution leaves in a 2D plane. For the B-sublattice component the vanishing of ${\bm\chi}_\Gamma(0)=0$ at the magic angle implies that solutions are collinear $\bm\chi_b\times\bm\chi_c=0$ and described by a Landau level like state with $C=1$.

\end{document}